\documentstyle[prl,aps,epsfig]{revtex}
\draft
\begin{document}
\twocolumn[
\hsize\textwidth\columnwidth\hsize\csname@twocolumnfalse\endcsname

\title{Kondo effect in a Luttinger liquid: nonuniversality of the Wilson
ratio}
\author{Karyn Le Hur\cite{New}}
\address{Laboratoire de Physique des Solides, Universit{\'e} Paris--Sud,
                    B{\^a}t. 510, 91405 Orsay, France}

\maketitle

\begin{abstract}
Using a coset
 Ising-Bose representation, we show how backscattering of electrons
off a magnetic impurity destabilizes the two-channel Kondo fixed point and
drives the system to a new fixed point, in agreement with previous 
results {\large \{}Phys. Rev. Lett. {\bf 75}, 300 (1995){\large \}}. In 
addition, we find that 
the presence of several leading 
correction-to-boundary-operators for nonzero U implies a nonuniversal Wilson 
ratio. Finally,
we show that a strong potential scattering at the impurity site stabilizes the
two-channel Kondo fixed point in the neighborhood of a half-filled band.

\end{abstract}

\vfill
 \twocolumn
\vskip.5pc ]
\narrowtext

The Kondo problem is one of the central topics in condensed matter
physics\cite{un}. A magnetic impurity embedded in a host metal may be the 
simplest
example of confinement, resulting from the growth of the effective Kondo
coupling $J_K$ at low temperatures\cite{deux}.
 The non-perturbative ground state 
is found to be of Fermi-liquid type\cite{trois,Andrei},
 where the quasiparticles acquire a
phase shift. The study of magnetic impurities in one-dimensional (1D) 
strongly correlated electron systems has particularly attracted great
interest in the last few years. Metals in
1D differ fundamentally from those in three 
dimensions, where the low
energy properties can be described very well by Landau's Fermi liquid theory. 
Electron-hole excitations are replaced by independent
charge and spin zero sound modes, leading to Luttinger
liquids (LL)\cite{cinq,quatre}.

The Kondo effect in a LL was recently considered by Furusaki and Nagaosa\cite
{huit},
expanding on earlier work by Lee and Toner\cite{sept}. 
Using poor man's scaling, an 
infinite-coupling
fixed point was identified, suggesting a complete screening of the impurity
at zero temperature. The authors clearly established that prominent
backscattering effects in a LL support a Kondo effect for ferromagnetic as 
well as antiferromagnetic Kondo
exchanges. The Kondo temperature yields a power-law dependence
on the exchange coupling, $T_K\propto {J_K}^{2/(1-K_{\rho})}$, $K_{\rho}$ being
the LL charge parameter. The impurity specific heat, as well as the
conductance, exhibit an anomalous temperature dependence, with a leading
term $T^{(1/K_{\rho})-1}$. But, it is unclear whether the
extrapolation of the scaling equations into the strong-coupling
regime is justified. Unfortunately, the Kondo model in a correlated 
host, where
the {\it dynamic} backscattering against the impurity crucially influences the
properties of the system, is nonintegrable. There are only few exact 
solutions with specific boundaries, based on new
Bethe-Ansatz solutions\cite{huit1,Zvyagin}.
\vskip 0.09cm
In this short paper we study the problem using boundary conformal field 
theory (BCFT). The heart of the method, pionnered by Affleck and Ludwig, is
to replace the impurity by a scale invariant boundary 
condition\cite{neuf,Ludwig}. Recent 
interesting BCFT
results by Fr\" ojdh and Johannesson\cite{onze} give only two scenario. 
Either the
system belongs to the Fermi liquid universality class or it indeed has the
properties predicted by Furusaki and Nagaosa. Several points, however, 
remained poorly understood. For instance, without backscattering effects
the fixed point corresponds to the
non-Fermi liquid of the standard two-channel Kondo
model\cite{douze}. It was unclear how the presence of the backward Kondo 
exchange definitely forbids such
a solution. We explain this point explicitly using
a precise Ising-Bose representation. Then, we 
show that the backscattering Kondo
exchange implies the presence of several leading 
correction-to-boundary-operators: the resulting Wilson 
ratio (of the Kondo effect in a LL) is {\it not}
well-defined despite the complete screening of
the impurity spin. Thus, the previous
Fermi liquid class\cite{onze,Schi} is not completely
universal in the critical region $T\ll T_K$. Finally, we show that 
a strong potential at 
the impurity site stabilizes the two-channel Kondo fixed point at
half-filling.

A key feature in the conformal-field-theory approach to this problem is 
{\it bosonization}\cite{cinq,neuf}. We begin to linearize the dispersion of 
conduction 
electrons; the lattice step is fixed to $a=1$. The relativistic fermions
$\psi_{\sigma}(x)$ are separated in left-movers $\psi_{L\sigma}(x)$ and
right-movers $\psi_{R\sigma}(x)$ on the Fermi-cone. Then, we introduce the
normal-ordered current operators for the charge and spin degrees of freedom,
namely $J_{L}=:\psi^{\dag}_{L\sigma}\psi_{L\sigma}(x):$ and ${\bf
J}_{L}=:\psi^{\dag}_{L\alpha}\sigma_{\alpha\beta}\psi_{L\beta}:$ and similarly
for the right-movers. To cast the problem on a form where BCFT applies, we
use a representation where the impurity location defines a boundary. For this
purpose we confine the system to the finite interval $x\in [-L,L]$ and define
right and left movers on the half plane $x\geq0$, with 
$\psi_{R\alpha}(t,x)\equiv
\psi_{L\alpha}(t,-x)$. Fields are only {\it left} movers and we can 
rename $\psi_{1\alpha}(x)=\psi_{L\alpha}(x)$,
$\psi_{2\alpha}(x)=\psi_{L\alpha}(-x)$, ${\bf
J}_1(x)={\bf J}_{L}(x)$ and ${\bf J}_2(x)={\bf J}_{L}(-x)$, and analogously 
for the charge sector.

The free spin theory of two channels of left-moving fermions can be
written in a Sugawara form. Let us introduce
important basic properties of Sugawara Hamiltonians. We start with a free 
spin Hamiltonian,
\begin{equation}
H_s=\frac{v_F}{2\pi}\int_{-L}^{L} dx\ {\cal H}_s(x)
\end{equation}
The Fermi velocity is $v_F=2t\sin k_F$. Assuming the boundary condition
${\cal H}_s(-L)={\cal H}_s(L)$, we can define 
Fourier modes generating the Virasoro algebra,
\begin{equation}
[L_n,L_m]=(n-m)L_{n+m}+\frac{C}{12}n(n^2-1)\delta_{n+m,0}
\end{equation}
where, $C$ is the conformal anomaly parameter. Considering low-dimensional
spin problems with SU(2) symmetry, it results a particular class of 
conformally invariant theories
which has a Hamiltonian density quadratic in the currents 
$J^a(x)$ $(a=1,2,3)$. The Fourier modes of the currents obey the Kac-Moody
algebra
\begin{equation}
[J_n^a,J_m^b]=i\epsilon^{abc}J_{n+m}^c+\frac{1}{2}k n\delta_{n+m,0}
\end{equation}
Here $k$ is the Kac-Moody level which must be a positive integer. The 
spin Hamiltonian then takes the Sugawara form
\begin{equation}
{\cal H}_s(x)=\frac{1}{2+k}: {\bf J}(x)\cdot {\bf J}(x):
\end{equation}
or, in momentum space, $L_n=\frac{1}{2+k}:{\bf J}_n{\bf J}_{n+m}:$.
Using the vacuum definition 
$\langle 0|[L_2,L_{-2}]|0\rangle=\frac{1}{2}C$, it follows that the 
Sugawara Hamiltonian obeys the Virasoro algebra with
conformal anomaly $C=3k/(2+k)$. We now return to the problem at hand. The free 
Hamiltonian is composed
of four commuting terms, quadratic in the charge and spin currents for
channels 1 and 2. The two sets of commuting spin currents for each channel
obey the Kac-Moody algebra at the level k=1, and the associated Hamiltonians,
$H_{si}$, each have a conformal anomaly $C=1$. It is natural to rewrite the 
theory in terms of the total current ${\bf J}={\bf J}_1+{\bf J}_2$. 
This obeys the Kac-Moody algebra with k=2. The associated Hamiltonian, $H_s$,  
constructed from ${\bf J}$ with k=2, has $C=3/2$. Thus, the Hamiltonian
$H_{s1}+H_{s2}$ can be written as a sum of two commuting pieces 
($H_s$) and a remainder
with a central charge $C=1/2$ (an Ising model). 

The former theory can be
considered to be the left-moving part of a k=2 Wess-Zumino-Witten (WZW) 
model. This is an example of the
Goddard-Kent-Olive coset construction\cite{GKO}. Considering the SU(2) Kac-
Moody
theory, there is a primary field of spin $s=0,\ 1/2,...,\ k/2$ with dimension
$\Delta=s(s+1)/(2+k)$. The k=2 WZW theory has primary of spin $s=0$ (identity
operator ${\mathbf 1}$), $s=1/2$ (fundamental field $g_{\alpha}$ of dimension
$\Delta=3/16$) and $s=1$ (denoted $\mathbf{\Phi}$
with $\Delta=1/2$). There are also three primary fields in the Ising model: the
identity operator $\mathbf{1}$, the Ising order parameter $\sigma$
with $\Delta=1/16$, and the energy operator $\epsilon$ with 
$\Delta=1/2$. To describe the charge degrees of freedom, we can 
simply introduce two bosons, $\phi_{c,i}$. In this representation the fermion 
field $(\alpha=\uparrow,\downarrow\ \hbox{and}\ j=1,2)$ is written as
\begin{equation}
\label{five}
\psi_{\alpha, j}\propto e^{\pm i\sqrt{2\pi}\phi_{c,j}}g_{\alpha}\sigma
\end{equation}
The fermion operator has the required dimension
$\Delta=1/4+3/16+1/16=1/2$. This type of representation is known as a
conformal embedding. We have the important {\it fusion rules},
\begin{equation}
\label{six}
g_{\alpha}\sigma_{\alpha}^{\beta}g_{\beta}\rightarrow\mathbf{1}+\Phi,\
\sigma\times\sigma\rightarrow \mathbf{1}+\epsilon
\end{equation}
The eigenstates of the {\it unperturbed} problem 
organize into a product of
two U(1) conformal towers labeled by the two quantum numbers $(Q,\Delta Q)$, 
respectively the {\it sum} and {\it
difference} of net charge between the two channels. The total charge $Q$ is
defined {\it modulo 2}. The complete set of
conformal tower is accordingly labeled  by $(Q,\Delta Q, s, \hbox{I})$ with
$\hbox{I}={\mathbf 1}, \sigma, \epsilon$. We introduce the total charge boson 
$\phi_c=\phi_{c1}+\phi_{c2}$ and the leftover degree of freedom
$\tilde{\phi}_c=\phi_{c1}-\phi_{c2}$ which describes
relative charge fluctuations. The free charge Hamiltonian can be identified 
as the well-known Luttinger model
\begin{equation}
 H_c=\frac{1}{2\pi}\int dx\
 \frac{u_{\rho}}{K_{\rho}}:{(\partial_x\phi_c)}^2:+u_{\rho}K_{\rho}:
 {(\pi\Pi_c)}^2 :
\end{equation}
Charge quasiparticles are holons. We have 
$\Pi_c(x)=\partial_x\tilde{\phi}_c(x)$ and $U$ describes the Hubbard 
interaction between electrons. The parameters $u_{\rho}$ and $K_{\rho}$ are 
given by $u_{\rho}K_{\rho}=v_F$
and $\frac{u_{\rho}}{K_{\rho}}=v_F+2U/\pi$. 
\vskip 0.06cm
It is crucial for the following to write the Kondo
interaction in the Bose-Ising representation. We obtain
\begin{equation}
\label{uno}
{\cal H}_{K}=[\lambda_f {\bf J}(0) +
\lambda_b\cos\sqrt{2\pi}\phi_c{\bf\Phi}(0)]{\bf S}
\end{equation} 
The terms $\lambda_f$ and $\lambda_b$, which are proportional
to $J_K$ refer to the forward 
and backward Kondo exchanges respectively. Both 
$\lambda_f$
and $\lambda_b$ flow to strong couplings when $T\rightarrow 0$\cite{huit}. 
For the 
derivation of the 
second term see, for instance, Part III of
ref.\cite{Jones}. Due to dimensional restrictions,
 the Ising sector is not coupled to the
impurity.
\vskip 0.06cm
First, let us briefly summarize the low-temperature properties when 
$\lambda_b=0$. We closely follow the notations of 
ref.\cite{neuf,Ludwig}. For the special value 
$\lambda_f^*=v_F/(2+k)|_{k=2}$, we can absorb 
the impurity spin by redefining the spin 
current as that of electrons and impurity, ${\bf J}(x)\rightarrow {\bf J}(x)+
2\pi{\bf S}\delta(x)$.
 Simply, this adds an extra
spin-$\frac{1}{2}$ degree of freedom to the $SU(2)_{k=2}$ spin sector. We
obtain the usual conformal fusion rules $s=0 \rightarrow \frac{1}{2}$,
$\frac{1}{2}\rightarrow 0\ or\ 1$, $1\rightarrow \frac{1}{2}$. It is
equivalent to a new boundary condition
$\psi_{1\alpha}(0)+\psi_{2\alpha}(0)=0$ which gives $\phi_c(0)=\sqrt{\pi/2}$
and then $\langle\cos\sqrt{2\pi}\phi_c(0)\rangle=$ constant. {\it The holons
move away from the impurity site}.
The special point $\lambda_f^*$ can
be identified as a strong coupling limit. At the impurity site, the 
interaction term will be the following configuration: one electron at the 
site 0 forms a singlet with the impurity $|\Uparrow\downarrow\rangle
-|\uparrow\Downarrow\rangle$ and another one remains 
{\it free} acting as an {\it effective} impurity. It describes the
 overscreened-like Kondo 
effect\cite{trois}. Curiously, the $\infty$-coupling is known to be not 
consistent in that case. The fixed point does not correspond to a simple
boundary condition on electrons, instead it is a Non-Fermi-Liquid fixed point. 
To show that, it is important to determine what 
the leading irrelevant operator (LIO) is 
which can appear in the effective theory around
 the {\it unstable} strong coupling fixed point. The LIO  have to respect the 
symmetry of the total Hamiltonian: they
must be SU(2) spin singlets, must respect the conservation of the
total charge Q and 
the channel interchange
symmetry $1\leftrightarrow 2$. To obtain an SU(2) singlet, other than the 
trivial identity operator, we can consider Kac-Moody descendants. The lowest 
dimension singlet is ${\bf J}_{-1}\cdot{\bf \Phi}$, of dimension 
$1+\Delta=1+2/(2+k)=3/2$
 where ${\bf \Phi}$ is the s=1 primary field\cite{neuf,Ludwig}. Adding 
$\delta H=\gamma_1{\bf J}_{-1}\cdot{\bf \Phi}$ to the effective Hamiltonian, 
the 
leading contribution to the specific heat and impurity susceptibility is second
order in $\gamma_1$:
\begin{equation}
C_{imp}=\gamma_1^2\pi^2 9T\ln\frac{T_K}{T},\ 
\chi_{imp}=18\gamma_1^2\ln\frac{T_K}{T}
\end{equation}
and a Wilson ratio,
\begin{equation}
R_W=\frac{\chi_{imp}}{C_{imp}}\frac{C}{\chi}=(1+K_{\rho})
\frac{(2+k/2)(2+k)^2}{36}|_{k=2}
\end{equation}
where, $C$ and $\chi$ are bulk quantities of the LL\cite{quatre}.
Simple scaling arguments give $\gamma_1\approx T_K^{-\frac{1}{2}}$.
For $U=0$ (or $K_{\rho}=1$), this reduces to the universal number 8/3 
characterizing the usual two-channel Kondo model. In the charge 
sector, all operators in a given conformal tower have the same charge; hence 
only 
descendants of the identity operator are permitted in the effective 
Hamiltonian.

Now, let us include backward scattering off the impurity, $\lambda_b=\lambda_f
\neq 0$. The important thing that we must do first, is to explain how 
backscattering of electrons off a magnetic impurity destabilizes the 
two-channel Kondo 
fixed point and drives the system to a new fixed point. First, 
${\bf J}_{-1}\cdot{\bf \Phi}$ is not allowed by {\it parity}. To obtain
a complete definition of parity, we take\cite{Jones}:
\begin{equation}
P_S: {\bf \Phi}\rightarrow -{\bf \Phi},\ {\bf J}_{-1}\rightarrow -{\bf J}_{-1},\ 
\phi_c\rightarrow -\phi_c
\end{equation}
Thus, it is easy to see that $\lambda_b$ is odd under parity whereas 
${\bf J}_{-1}\cdot{\bf \Phi}$ is even. As in the two-impurity case, such an 
operator is absent by parity\cite{Jones}. Second, using 
the Bose representation (\ref{uno}), we can obtain more complete 
results. Indeed, 
when $\lambda_b$ goes to {\it infinity}, we deduce that
any operator coming from the s=1 tower {\it must}
transform as triplets in the neighborood of the impurity
site. In consequence, the expectation value of
${\bf J}_{-1}\cdot{\bf \Phi}$ (which is a s=1-spin singlet)
becomes equal to {\it zero} by this
new boundary condition at 
the origin. No LIO can now arise from the s=1 tower.
It can also be useful to rewrite the problem in the original (1,2)
basis. It may be simply performed using the common identifications: ${\bf
  J}_{-1}\rightarrow{\bf J}^{-1}_1+{\bf J}^{-1}_2,\
{\bf\Phi}\rightarrow\hbox{Tr}(h{\pmb{$\tau$}})$ where
$h\in SU(2)_{k=1}$ and ${\pmb{$\tau$}}_i$ are Pauli matrices,
\begin{equation}
{\bf J}_{-1}\cdot{\bf \Phi}(0)=({\bf J}^{-1}_1+{\bf J}^{-1}_2)
\hbox{Tr}(h{\pmb{$\tau$}})(0)=\frac{d}{dx}\hbox{Tr}h(0)
\end{equation}
Starting with an Heisenberg chain with S=1/2, such an operator corresponds to
vary two adjacent links by the same amount\cite{onzei}. When 
$\lambda_b=0$, we may think of the 
electrons (one
from each channel) in the first layer around the impurity as aligning 
antiferromagnetically with the impurity. This overscreens it, leaving 
an effective impurity. Then, the electrons in the next layer overscreens 
this effective impurity through a coupling $\lambda_{eff}\sim t^2/J_K$, 
destabilizing the 
strong coupling fixed point. Through virtual excitations of the two
competing singlet states, attraction can
be mediated between the two electrons in the first layer and 
that of the second one which participates in the
Kondo screening of the effective impurity. This 
requires the same LIO as for the Kondo problem in
a {\it periodic} Heisenberg chain. At each stage, we have an effective s=1/2 
impurity. There is a symmetry between strong- and weak-coupling 
regions. When $\lambda_b\neq 0$, this picture breaks down because 
exchange 
processes between the two channels (via the impurity) 
are prominent at the origin. 
 Thus, the Kondo screening should involve
 the two electronic channels in a symmetric
way. It
stabilizes the strong coupling regime and destabilizes the usual 
two-channel Kondo fixed point. {\it It gives a new fixed point} in accordance 
with ref.\cite{onze}. We expect two cases. 
\vskip 0.18cm
First, the exchange term $\lambda_b$ breaks the U(1) invariance. Thus, 
$\Delta Q$ is no longer restricted to zero, and the charge
sector makes nontrivial contributions to the content of scaling operators. The
lowest dimension operator allowed by the forward scattering selection rule is
obtained from $(\Delta Q,Q)=(\pm 2,0)$. The sector $\Delta Q=\pm 2$ is
described through the term $\psi_{\alpha,1}^{\dag}\psi_{\beta,2}(0)$, where
$\phi_c\rightarrow \tilde{\phi}_c$. Unlike in Fermi liquids, it can produce
exotic tunneling phenomena in Luttinger liquids\cite{six}. Using Eq.
(\ref{five}) and choosing the set $(s,\hbox{I})=({\mathbf 1},\epsilon)$
in the fusion rules (\ref{six}), we obtain the
{\it lowest} dimension LIO,
$\gamma_{2}\cos\sqrt{2\pi}\tilde{\phi}_c(0)\cdot\epsilon$. Such an
 operator has 
the scaling dimension
$\Delta=\frac{1}{2}(\frac{1}{K_{\rho}}+1)$ and then $\gamma_{2}\approx
1/\lambda_b$. Note that
a tunneling
process is obtained by coupling 
$\cos\sqrt{2\pi}\tilde{\phi}_c(0)$ and the Ising sector (the spin triplet
$\Phi$ is coupled to the impurity). 
Since $\epsilon$ fusion
takes spin up into spin down and vice versa, $\gamma_{2}$ describes tunneling
processes with spin flip, which break the spin singlet at 
the impurity site.
When $T\rightarrow 0$, the breakage of the singlet
becomes impossible because the strong Kondo coupling fixed point is 
consistent. Nevertheless, $\gamma_{2}$ affects the low-energy behavior. The 
leading contribution to the specific heat is second order in $\gamma_2$. Using
simple scaling arguments, we obtain $C_{tun}\propto
\gamma_{2}^2 T^{(1/K_{\rho})-1}$. We find that
the susceptibility is not corrected by processes in 
${\cal O}(\gamma_{2}^2)$. The term $\gamma_2$ also modifies the universal
 conductance of the LL, $G_o=2e^2K_{\rho}/h$ obtained by applying a 
static field over a finite part of the sample\cite{quatre}. 
We can define the (linear response) conductance as
$G_{tun}(T)\propto\gamma_2(T)^2$\cite{six}. Using the $\beta$-function of 
$\gamma_{2}$: $d\gamma_2/d\ln T=\frac{1}{2}(\frac{1}{K_{\rho}}-1)\gamma_2$,
we find that the conductance varies as $G_{tun}(T)\sim 
\gamma_{2}^2 T^{(1/K_{\rho})-1}$. Physical properties exhibit
an exact duality between high- and
low-temperature fixed points under replacement $\lambda_b\rightarrow \gamma_2$,
$K_{\rho}\rightarrow 1/K_{\rho}$\cite{Lesage}.
This scaling agrees exactly with that 
in ref.\cite{huit}.

Second, we must carefully re-investigate the $SU(2)_2$ sector. In fact, the 
leading operator $\gamma_{3}{\bf J}(x){\bf J}(x)\delta(x)$ may be 
involved in the strong coupling physics because $\gamma_1=0$. When
$\gamma_1\neq 0$, such an operator is expelled because its dimension yields
 $2>3/2$. Since the two channels both participate in the impurity screening, 
there are {\it three} leading irrelevant operators, all of 
dimension 2: ${\bf J}_{1}^2$, ${\bf J}_2^2$, ${\bf J}_{1}{\bf J}_2$. 
Note that the complex screening of the impurity occurs over a large
length scale, $\xi_K=\hbar v_F/k_B T_K$ so that interactions between
channels 1 and 2 are assured. We expect the corresponding coupling 
constants to
be of order $1/T_K$. The impurity specific heat $C_{imp}$ is proportional to
$T/T_K$, since it arises from first-order perturbation theory in the leading 
irrelevant operators. By the same reasoning, we obtain $\chi_{imp}$ 
proportional to $1/T_K$, T-independent when $T\rightarrow 0$. 
We close this 
part with an important discussion about the nonuniversality of the Wilson 
ratio $R_W$ in that case. Wilson\cite{deux} and Nozi\`eres\cite{trois}
 stressed that, for the 
exactly screened case the
Wilson ratio is always universal in the weak-coupling limit. Starting with a 
magnetic impurity in a purely
1D interacting system, we have found three leading irrelevant 
operators: $R_W$ is not {\it universal} despite the complete
screening of the impurity. 
The nonuniversality in the Wilson ratio in the spinon basis
produces a 
low-temperature behavior which is not completely determined: 
a `pathological' local Fermi-liquid fixed point could also take place.
Note that we do not predict many drastic changes by approaching 
half-filling. The so-called Umklapp
processes produce a charge gap. The ultraviolet cut-off is rescaled
as $T_c\propto U^{1/2(1-K_{\rho})}$\cite{quatre}. At low temperature,
$U\rightarrow +\infty$ and the system becomes 
{\it insulating}. The Kondo energy scale is of course affected; we find:  
$T_K\propto T_c\lambda_b^2$. For $T\ll T_K$, we have still
$\lambda_f=\lambda_b\rightarrow +\infty$ and then we predict the same
fixed point as before.  We can note that whatever the filling,
the system has the same ground state
 as an {\it open} Heisenberg chain due to backscattering\cite{onzei}.

Finally, we re-discuss the metal-insulator transition when adding
a {\it strong} scattering potential, $V\rightarrow +\infty$ at the 
origin. Away from half-filling, it separates the problem 
into two semi-infinite Luttinger liquids. The term $\lambda_b$
must be carefully replaced by,
\begin{equation}
\lambda_b{\bf\Phi}(0).\cos(\sqrt{2\pi}\tilde{\phi}_c)(0){\bf S}
\end{equation}
It generates tunneling processes with flip of the
impurity spin; the bare condition is
fixed to: $\lambda_b\sim 1/V$. Using an operator product expansion, we predict:
\begin{equation}
\frac{d\lambda_b}{d\ln L}=\frac{1}{2\pi v_F}(2\lambda_f-U)\lambda_b
\end{equation}
Starting with $2\lambda_f<U$, a 
 small $\lambda_b$ will firstly decrease. On the
other hand, $\lambda_f$ is a marginal 
interaction and flows to strong couplings at 
$T_K\propto e^{-2\pi v_F/\lambda_f}$. The increase of 
$\lambda_f$ will ultimately drive $\lambda_b$ to
strong values as soon as $\lambda_f(T)>U/2$. The
presence of a strong potential scattering cannot suppress backscattering of
electrons off the magnetic impurity, in accordance with results of
ref.\cite{fab}. In contrast, an insulating state with
$U\rightarrow +\infty$ can freeze any exchange process between the
two channels, and
 $\lambda_b\rightarrow 0$ whatever the renormalization of
 $\lambda_f$. At half-filling, the two-channel Kondo fixed
point is the unique solution. For strong $V$, we pass 
from a periodic 
to an open Heisenberg chain by doping.

To conclude, the Ising-Bose representation shows accurately how
backscattering of electrons off a magnetic impurity destabilizes the
two-channel Kondo fixed point, verifies 
the scaling proposed by Furusaki
and Nagaosa, and proves that the previous local Fermi liquid class is
not completely universal
 when backscattering is included because the Wilson ratio
is not well-defined.

\end{document}